\documentclass[pre,aps,amssymb,amsmath,superscriptaddress,twocolumn]{revtex4}

\usepackage{graphicx}
\usepackage{dcolumn}
\usepackage{bm}
\usepackage{epstopdf}

\newcount\driver
\newcount\bozza

\font\cs=cmcsc10 scaled\magstep1
\font\ottorm=cmr8 scaled\magstep1 \font\msxtw=msbm10
scaled\magstep1                  \font\euftw=eufm10
scaled\magstep1 \font\msytw=msbm10 scaled\magstep1
\font\msytww=msbm8 scaled\magstep1 \font\msytwww=msbm7
scaled\magstep1                 \font\indbf=cmbx10 scaled\magstep2
\font\grbold=cmmib10 scaled\magstep1
\font\amit=cmmi7 \def\sf{\textfont1=\amit} \font\bigtenrm=cmr10
scaled \magstep2               \font\bigteni=cmmi10 scaled
{\count255=\time\divide\count255 by 60
\xdef\hourmin{\number\count255}
        \multiply\count255 by-60\advance\count255 by\time
   \xdef\hourmin{\hourmin:\ifnum\count255<10 0\fi\the\count255}}

\let\a=\alpha \let\b=\beta    \let\g=\gamma     \let\d=\delta     \let\e=\varepsilon
\let\z=\zeta  \let\h=\eta     \let\th=\vartheta \let\k=\kappa     \let\l=\lambda
\let\m=\mu    \let\n=\nu      \let\x=\xi        \let\p=\pi        \let\r=\rho
\let\s=\sigma \let\t=\tau     \let\f=\varphi    \let\ph=\varphi   \let\c=\chi
\let\ps=\psi  \let\y=\upsilon \let\o=\omega     \let\si=\varsigma
\let\G=\Gamma \let\D=\Delta   \let\Th=\Theta    \let\L=\Lambda    \let\X=\Xi
\let\P=\Pi    \let\Si=\Sigma  \let\F=\Phi       \let\Ps=\Psi
\let\O=\Omega \let\Y=\Upsilon

\def\PP{{\cal P}}\def\EE{{\cal E}}\def\MM{{\cal M}}\def\VV{{\cal V}}
\def\FF{{\cal F}}\def\HH{{\cal H}}\def\WW{{\cal W}}
\def\TT{{\cal T}}\def\NN{{\cal N}}\def\BB{{\cal B}}\def\ZZ{{\cal Z}}
\def\RR{{\cal R}}\def\LL{{\cal L}}\def\JJ{{\cal J}}\def\QQ{{\cal Q}}
\def\DD{{\cal D}}\def\AA{{\cal A}}\def\GG{{\cal G}}\def\SS{{\cal S}}
\def\OO{{\cal O}}\def\XXX{{\bf X}}\def\YYY{{\bf Y}}\def\WWW{{\bf W}}
\def\KK{{\cal K}}

\def\pp{{\bf p}}\def\qq{{\bf q}}\def\ii{{\bf i}}\def\xx{{\bf x}}
\def\aaa{{\bf a}} \def\bb{{\bf b}} \def\dd{{\bf d}}
\def\yy{{\bf y}}\def\kk{{\bf k}}\def\mm{{\bf m}}\def\nn{{\bf n}}
\def\zz{{\bf z}}\def\uu{{\bf u}}\def\vv{{\bf v}}\def\ww{{\bf w}}
\def\xxi{\hbox{\grbold \char24}} \def\bP{{\bf P}}\def\rr{{\bf r}}
\def\tt{{\bf t}}\def\bT{{\bf T}}

\def\ss{{\underline \sigma}}       \def\oo{{\underline \omega}}
\def\ee{{\underline \varepsilon}}  \def\aa{{\underline \alpha}}
\def\un{{\underline \nu}}          \def\ul{{\underline \lambda}}
\def\um{{\underline \mu}}          \def\ux{{\underline\xx}}
\def\uk{{\underline \kk}}          \def\uq{{\underline\qq}}
\def\uaa{{\underline \aaa}} \def\ub{{\underline\bb}}
\def\uc{{\underlinec}} \def\ud{{\underline\dd}}
\def\up{{\underline\pp}}           \def\ua{{\underline \a}}
\def\ut{{\underline t}}            \def\uxi{{\underline \xi}}
\def\umu{{\underline \m}}          \def\uv{{\underline\vv}}
\def\ue{{\underline \e}}           \def\uy{{\underline\yy}}
\def\uz{{\underline \zz}}
\def\uw{{\underline \ww}}          \def\uo{{\underline \o}}
\def\us{{\underline \s}}           \def\xxx{{\underline \xx}}
\def\kkk{{\underline\kk}}          \def\uuu{{\underline\uu}}
\def\udpr{{\underline\Dpr}}

\def\uu{\bf u}
\def\III{\hbox{\msytw I}}
\def\MMM{\hbox{\euftw M}}          \def\BBB{\hbox{\euftw B}}
\def\RRR{\hbox{\msytw R}}          \def\rrrr{\hbox{\msytww R}}
\def\rrr{\hbox{\msytwww R}}        

\def\NNN{\hbox{\msytw N}}          \def\nnnn{\hbox{\msytww N}}
\def\nnn{\hbox{\msytwww N}}        \def\ZZZ{\hbox{\msytw Z}}
\def\zzzz{\hbox{\msytww Z}}        \def\zzz{\hbox{\msytwww Z}}
\def\TTT{\hbox{\msytw T}}          \def\tttt{\hbox{\msytww T}}
\def\ttt{\hbox{\msytwww T}}        \def\EE{\hbox{\msytw E}}
\def\eeee{\hbox{\msytww E}}        \def\eee{\hbox{\msytwww E}}

\let\dpr=\partial
\let\circa=\cong
\let\bs=\backslash
\let\==\equiv
\let\txt=\textstyle
\let\io=\infty
\let\0=\noindent

\def\pagina{{\vfill\eject}}
\def\*{{\hfill\break\null\hfill\break}}
\def\bra#1{{\langle#1|}}
\def\ket#1{{|#1\rangle}}
\def\media#1{{\langle#1\rangle}}
\def\ie{\hbox{\it i.e.\ }}
\def\eg{\hbox{\it e.g.\ }}

\def\tilde#1{{\widetilde #1}}

\def\Dpr{\V\dpr\,}
\def\aps{{\it a posteriori}}
\def\lft{\left}
\def\rgt{\right}
\def\der{\hbox{\rm d}}
\def\la{{\langle}}
\def\ra{{\rangle}}
\def\norm#1{{\left|\hskip-.05em\left|#1\right|\hskip-.05em\right|}}
\def\tgl#1{\!\!\not\!#1\hskip1pt}
\def\tende#1{\,\vtop{\ialign{##\crcr\rightarrowfill\crcr
             \noalign{\kern-1pt\nointerlineskip}
             \hskip3.pt${\scriptstyle #1}$\hskip3.pt\crcr}}\,}
\def\otto{\,{\kern-1.truept\leftarrow\kern-5.truept\to\kern-1.truept}\,}
\def\fra#1#2{{#1\over#2}}

\def\sde{{\cs SDe}}
\def\wti{{\cs WTi}}
\def\osa{{\cs OSa}}
\def\ce{{\cs CE}}
\def\rg{{\cs RG}}

\def\lp{{\hskip-1pt:\hskip 0pt}}
\def\rp{{\hskip-1pt :\hskip1pt}}
\def\defi{{\buildrel \;def\; \over =}}
\def\apt{{\;\buildrel apt \over =}\;}
\def\nequiv{\not\equiv}
\def\Tr{\rm Tr}
\def\diam{{\rm diam}}
\def\sgn{\rm sgn}
\def\wt#1{\widetilde{#1}}
\def\wh#1{\widehat{#1}}
\def\hat#1{\wh{#1}}
\def\sqt[#1]#2{\root #1\of {#2}}

\def\ha{{\widehat \a}}\def\hx{{\widehat \x}}\def\hb{{\widehat \b}}
\def\hr{{\widehat \r}}\def\hw{{\widehat w}}\def\hv{{\widehat v}}
\def\hf{{\widehat \f}}\def\hW{{\widehat W}}\def\hH{{\widehat H}}
\def\hB{{\widehat B}}
\def\hK{{\widehat K}} \def\hW{{\widehat W}}\def\hU{{\widehat U}}
\def\hp{{\widehat \ps}}  \def\hF{{\widehat F}}
\def\bp{{\bar \ps}}
\def\hh{{\hat \h}}
\def\jm{{\jmath}}
\def\hJ{{\widehat \jmath}}
\def\hJ{{\widehat J}}
\def\hg{{\widehat g}}
\def\tg{{\tilde g}}
\def\hQ{{\widehat Q}}
\def\hC{{\widehat C}}
\def\hA{{\widehat A}}
\def\hD{{\widehat \D}}
\def\hDD{{\hat \D}}
\def\bl{{\bar \l}}
\def\hG{{\widehat G}}
\def\hS{{\widehat S}}
\def\hR{{\widehat R}}
\def\hM{{\widehat M}}
\def\hN{{\widehat N}}
\def\hn{{\widehat \n}}

\def\PP{{\cal P}}\def\EE{{\cal E}}\def\MM{{\cal M}}\def\VV{{\cal V}}
\def\FF{{\cal F}}\def\HH{{\cal H}}\def\WW{{\cal W}}
\def\TT{{\cal T}}\def\NN{{\cal N}}\def\BB{{\cal B}}\def\ZZ{{\cal Z}}
\def\RR{{\cal R}}\def\LL{{\cal L}}\def\JJ{{\cal J}}\def\QQ{{\cal Q}}
\def\DD{{\cal D}}\def\AA{{\cal A}}\def\GG{{\cal G}}\def\SS{{\cal S}}
\def\OO{{\cal O}}\def\AAA{{\cal A}}

\def\T#1{{#1_{\kern-3pt\lower7pt\hbox{$\widetilde{}$}}\kern3pt}}
\def\VVV#1{{\underline #1}_{\kern-3pt
\lower7pt\hbox{$\widetilde{}$}}\kern3pt\,}
\def\W#1{#1_{\kern-3pt\lower7.5pt\hbox{$\widetilde{}$}}\kern2pt\,}
\def\Re{{\rm Re}\,}\def\Im{{\rm Im}\,}
\def\lis{\overline}\def\tto{\Rightarrow}
\def\etc{{\it etc}} \def\acapo{\hfill\break}
\def\per{{\rm per}\,} \def\sign{{\rm sign}\,}
\def\indica{\leaders \hbox to 0.5cm{\hss.\hss}\hfill}
\def\guida{\leaders\hbox to 1em{\hss.\hss}\hfill}
\mathchardef\oo= "0521

\def\V#1{{\bf #1}}
\def\pp{{\bf p}}\def\qq{{\bf q}}\def\ii{{\bf i}}\def\xx{{\bf x}}
\def\yy{{\bf y}}\def\kk{{\bf k}}\def\mm{{\bf m}}\def\nn{{\bf n}}
\def\dd{{\bf d}}\def\zz{{\bf z}}\def\uu{{\bf u}}\def\vv{{\bf v}}
\def\xxi{\hbox{\grbold \char24}} \def\bP{{\bf P}}\def\rr{{\bf r}}
\def\tt{{\bf t}} \def\bz{{\bf 0}}
\def\ss{{\underline \sigma}}\def\oo{{\underline \omega}}
\def\xxx{{\underline\xx}}
\let\ciao=\bye
\def\qed{\raise1pt\hbox{\vrule height5pt width5pt depth0pt}}
\def\barf#1{{\tilde \f_{#1}}} \def\tg#1{{\tilde g_{#1}}}
\def\bq{{\bar q}} \def\bh{{\bar h}} \def\bp{{\bar p}} \def\bpp{{\bar \pp}}
\def\Val{{\rm Val}}
\def\indic{\hbox{\raise-2pt \hbox{\indbf 1}}}
\def\bk#1#2{\bar\kk_{#1#2}}
\def\tdh{{\tilde h}}

\def\RRR{\hbox{\msytw R}} \def\rrrr{\hbox{\msytww R}}
\def\rrr{\hbox{\msytwww R}} 
\def\NNN{\hbox{\msytw N}} \def\nnnn{\hbox{\msytww N}}
\def\nnn{\hbox{\msytwww N}} \def\ZZZ{\hbox{\msytw Z}}
\def\zzzz{\hbox{\msytww Z}} \def\zzz{\hbox{\msytwww Z}}
\def\TTT{\hbox{\msytw T}} \def\tttt{\hbox{\msytww T}}
\def\ttt{\hbox{\msytwww T}}

\def\ins#1#2#3{\vbox to0pt{\kern-#2 \hbox{\kern#1 #3}\vss}\nointerlineskip}

\newdimen\xshift \newdimen\xwidth \newdimen\yshift

\def\insertplot#1#2#3#4#5#6{%
\xwidth=#1pt \xshift=\hsize \advance\xshift by-\xwidth \divide\xshift by 2%
\begin{figure}[ht]
\vspace{#2pt} \hspace{\xshift}
\begin{minipage}{#1pt}
#3 \ifnum\driver=1 \griglia=#6
\ifnum\griglia=1 \openout13=griglia.ps \write13{gsave .2
setlinewidth} \write13{0 10 #1 {dup 0 moveto #2 lineto } for}
\write13{0 10 #2 {dup 0 exch moveto #1 exch lineto } for}
\write13{stroke} \write13{.5 setlinewidth} \write13{0 50 #1 {dup 0
moveto #2 lineto } for} \write13{0 50 #2 {dup 0 exch moveto #1
exch lineto } for} \write13{stroke grestore} \closeout13
\includegraphics{griglia.ps} \fi
\includegraphics{#4.ps}\fi%
\ifnum\driver=2 \fi
\end{minipage}
\caption{#5}
\end{figure}
}

\def\gtopl{\hbox{\msxtw \char63}}
\def\ltopg{\hbox{\msxtw \char55}}

\newdimen\shift \shift=-1.5truecm
\def\lb#1{%
\ifnum\bozza=1
\label{#1}\rlap{\hbox{\hskip\shift$\scriptstyle#1$}}
\else\label{#1} \fi}

\def\be{\begin{equation}}
\def\ee{\end{equation}}
\def\bea{\begin{eqnarray}}\def\eea{\end{eqnarray}}
\def\bean{\begin{eqnarray*}}\def\eean{\end{eqnarray*}}
\def\bfr{\begin{flushright}}\def\efr{\end{flushright}}
\def\bc{\begin{center}}\def\ec{\end{center}}
\def\bal{\begin{align}}\def\eal{\end{align}}
\def\ba#1{\begin{array}{#1}} \def\ea{\end{array}}
\def\bd{\begin{description}}\def\ed{\end{description}}
\def\bv{\begin{verbatim}}\def\ev{\end{verbatim}}
\def\nn{\nonumber}
\def\Halmos{\hfill\vrule height10pt width4pt depth2pt \par\hbox to \hsize{}}
\def\pref#1{(\ref{#1})}
\def\Dim{{\bf Dim. -\ \ }} \def\Sol{{\bf Soluzione -\ \ }}
\def\virg{\quad,\quad}
\def\bsl{$\backslash$}

\def\ins#1#2#3{\vbox to0pt{\kern-#2 \hbox{\kern#1 #3}\vss}\nointerlineskip}

\newdimen\xshift \newdimen\xwidth \newdimen\yshift
\newcount\griglia

\def\insertplot#1#2#3#4#5#6{%
\xwidth=#1pt \xshift=\hsize \advance\xshift by-\xwidth \divide\xshift by 2%
\begin{figure}[ht]
\vspace{#2pt} \hspace{\xshift}
\begin{minipage}{#1pt}
#3 \ifnum\driver=1 \griglia=#6
\ifnum\griglia=1 \openout13=griglia.ps \write13{gsave .2
setlinewidth} \write13{0 10 #1 {dup 0 moveto #2 lineto } for}
\write13{0 10 #2 {dup 0 exch moveto #1 exch lineto } for}
\write13{stroke} \write13{.5 setlinewidth} \write13{0 50 #1 {dup 0
moveto #2 lineto } for} \write13{0 50 #2 {dup 0 exch moveto #1
exch lineto } for} \write13{stroke grestore} \closeout13
\includegraphics{griglia.ps} \fi
\includegraphics{#4.ps}\fi%
\ifnum\driver=2 \fi
\end{minipage}
\caption{#5}
\end{figure}
}

\def\gtopl{\hbox{\msxtw \char63}}
\def\ltopg{\hbox{\msxtw \char55}}

\newdimen\shift \shift=-1.5truecm
\def\lb#1{%
\label{#1}\rlap{\hbox{\hskip\shift$\scriptstyle#1$}}
\else\label{#1} \fi}

\def\be{\begin{equation}}
\def\ee{\end{equation}}
\def\bea{\begin{eqnarray}}\def\eea{\end{eqnarray}}
\def\bean{\begin{eqnarray*}}\def\eean{\end{eqnarray*}}
\def\bfr{\begin{flushright}}\def\efr{\end{flushright}}
\def\bc{\begin{center}}\def\ec{\end{center}}
\def\bal{\begin{align}}\def\eal{\end{align}}
\def\ba#1{\begin{array}{#1}} \def\ea{\end{array}}
\def\bd{\begin{description}}\def\ed{\end{description}}
\def\bv{\begin{verbatim}}\def\ev{\end{verbatim}}
\def\nn{\nonumber}
\def\Halmos{\hfill\vrule height10pt width4pt depth2pt \par\hbox to \hsize{}}
\def\pref#1{(\ref{#1})}
\def\Dim{{\bf Dim. -\ \ }} \def\Sol{{\bf Soluzione -\ \ }}
\def\virg{\quad,\quad}
\def\bsl{$\backslash$}

\usepackage{amsmath}
\usepackage{amsfonts}
\usepackage{amssymb}
\usepackage{epstopdf}

\font\msytw=msbm9 scaled\magstep1 \font\msytww=msbm7
scaled\magstep1 \font\msytwww=msbm5 scaled\magstep1
\font\cs=cmcsc10

\let\a=\alpha \let\b=\beta  \let\g=\gamma  \let\d=\delta
\let\e=\varepsilon
\let\z=\zeta  \let\h=\eta   \let\th=\theta \let\k=\kappa \let\l=\lambda
\let\m=\mu    \let\n=\nu    \let\x=\xi     \let\p=\pi    \let\r=\rho
\let\s=\sigma \let\t=\tau   \let\f=\varphi \let\ph=\varphi\let\c=\chi
\let\ps=\Psi  \let\y=\upsilon \let\o=\omega\let\si=\varsigma
\let\G=\Gamma \let\D=\Delta  \let\Th=\Theta\let\L=\Lambda \let\X=\Xi
\let\P=\Pi    \let\Si=\Sigma \let\F=\Phi    \let\Ps=\Psi
\let\O=\Omega \let\Y=\Upsilon

\def\PPP{{\cal P}}\def\EE{{\cal E}}\def\MM{{\cal M}} \def\VV{{\cal V}}
\def\FF{{\cal F}} \def\HHH{{\cal H}}\def\WW{{\cal W}}
\def\TT{{\cal T}}\def\NN{{\cal N}} \def\BBB{{\cal B}}\def\III{{\cal I}}
\def\RR{{\cal R}}\def\LL{{\cal L}} \def\JJ{{\cal J}} \def\OO{{\cal O}}
\def\DD{{\cal D}}\def\AAA{{\cal A}}\def\GG{{\cal G}} \def\SS{{\cal S}}
\def\KK{{\cal K}}\def\UU{{\cal U}} \def\QQ{{\cal Q}} \def\XXX{{\cal X}}

\def\qq{{\bf q}} \def\pp{{\bf p}}
\def\vv{{\bf v}} \def\xx{{\bf x}} \def\yy{{\bf y}} \def\zz{{\bf z}}
\def\aa{{\bf a}}\def\hh{{\bf h}}\def\kk{{\bf k}}
\def\mm{{\bf m}}\def\PP{{\bf P}}

\def\dd{{\boldsymbol{\delta}}}

\def\ddd{\boldsymbol{\d}}
\def\TTTT{\mathbf{T}}

\def\nn{\nonumber}
\def\us{\underset}
\def\os{\overset}

\def\RRR{\hbox{\msytw R}} \def\rrrr{\hbox{\msytww R}}
\def\rrr{\hbox{\msytwww R}}
\def\NNN{\hbox{\msytw N}} \def\nnnn{\hbox{\msytww N}}
\def\nnn{\hbox{\msytwww N}} \def\ZZZ{\hbox{\msytw Z}}
\def\zzzz{\hbox{\msytww Z}} \def\zzz{\hbox{\msytwww Z}}
\def\TTT{\hbox{\msytw T}}

\def\\{\hfill\break}
\def\={:=}
\let\io=\infty
\let\0=\noindent\def\pagina{{\vfill\eject}}
\def\media#1{{\langle#1\rangle}}
\let\dpr=\partial
\def\sign{{\rm sign}}
\def\const{{\rm const}}
\def\tende#1{\,\vtop{\ialign{##\crcr\rightarrowfill\crcr\noalign{\kern-1pt
    \nointerlineskip} \hskip3.pt${\scriptstyle #1}$\hskip3.pt\crcr}}\,}
\def\otto{\,{\kern-1.truept\leftarrow\kern-5.truept\to\kern-1.truept}\,}
\def\defin{{\buildrel def\over=}}
\def\wt{\widetilde}
\def\wh{\widehat}
\def\to{\rightarrow}
\def\la{\left\langle}
\def\ra{\right\rangle}
\def\qed{\hfill\raise1pt\hbox{\vrule height5pt width5pt depth0pt}}
\def\Val{{\rm Val}}
\def\ul#1{{\underline#1}}
\def\lis{\overline}
\def\V#1{{\bf#1}}
\def\be{\begin{equation}}
\def\ee{\end{equation}}
\def\bp{\begin{pmatrix}}
\def\ep{\end{pmatrix}}
\def\bea{\begin{eqnarray}}
\def\eea{\end{eqnarray}}
\def\nn{\nonumber}
\def\pref#1{(\ref{#1})}
\def\ie{{\it i.e.}}
\def\lb{\label}
\def\eg{{\it e.g.}}

\def\Tr{\mathrm{Tr}}
\def\eu{\mathrm{e}}

\begin{document}

\title{Interacting
spinning fermions with quasi-random disorder}

\author{Vieri Mastropietro}

\address{
Universit\'a di Milano, Via C. Saldini 50, 20133, Milano, Italy
}

\begin{abstract} 
Interacting spinning fermions with strong quasi-random disorder are analyzed via
rigorous Renormalization Group (RG) methods combined with KAM techniques. The correlations are written in terms of an expansion whose convergence follows from 
number-theoretical properties of the frequency and cancellations due to Pauli principle.
A striking difference appears between spinless and spinning fermions; in the first case there are no relevant effective interactions while 
in presence of spin an additional relevant 
quartic term is present in the RG flow. 
The large distance exponential decay of the correlations present in the non interacting case, consequence of the single particle  localization, is shown to persist
in the spinning case 
only for temperatures greater than a power of the many body interaction, while in the spinless case this happens up to zero temperature.
\end{abstract}


\maketitle

\renewcommand{\thesection}{\arabic{section}}

Disorder can produce localization of quantum particles
\cite{A}, whose persistence 
in presence of interaction is known as
{\it many body localization} (MBL)  \cite{PH1},\cite{PH4}. 
Even if interacting,
systems displaying MBL are not expected to reach thermal equilibrium and consequently
their long-time properties 
are not captured by the conventional ensembles of quantum statistical mechanics, see e.g. \cite{NH},\cite{HL}.  
Theoretically establishing the MBL phase requires to rule
out possible nonperturbative effects which may cause delocalization and is therefore a subtle problem.
First analysis on the interplay of localization and interaction focused on ground state properties
\cite{FA},\cite{GGG}; in more recent times 
evidence for MBL was found in
\cite{Ba} (see also \cite{Ba1},\cite{Ba2}),
by constructing a complete set of conserved quasi-
local quantities in terms of formal series. Even if
defined order by order in perturbation theory, such series could diverge and MBL could only be metastable
 , see \cite{Ro}, as it happens in classical mechanics
in Birkhoff series, see e.g. \cite{Ga}, due to a phenomenon called
small divisor problem. An important progress toward a full
non-perturbative proof of MBL has been recently obtained in \cite{loc4}, \cite{loc4a}
but the proof is based on an assumption called
{\it limited level attraction} (eigenvalues do not accumulate too strongly near 
states with almost the same energies) which is still unproven. 
Other proofs of MBL are devoted to systems 
which can be mapped in non interacting ones \cite{S1},\cite{S2}.

In recent cold atoms experiments \cite{B} evidence of MBL has been reported.  The disorder is not 
random  (as in the above mentioned theoretical works)
but {\it quasi-random}, but the theory of MBL can be developed also in that case \cite{H4}. The experiments
use two laser beams with incommensurate frequencies,
superimposing to a one dimensional lattice a periodic potential 
with a period that is incommensurate with  the  underlying  lattice, producing a realization
of an interacting {\it 
Aubry-Andre'} model \cite{AA}; subsequent experiments considered two coupled chains \cite{B1}. 

Random or quasi random disorder have similar properties, at least for strong disorder. In particular, a strong incommensurate potential  produces localization of the single particles eigenstates
\cite{FS},\cite{Ia}, as in the random case, while for weak potential there is no localization. There is then a metal-insulator transition as for 3D random disorder (in 1D the eigenstates in the random case are instead always localized). Such properties are due to a close relation with 
the Kolmogorov-Arnold-Moser (KAM) theorem for the stability of tori of perturbed Hamiltonian systems. The simplest 
generalization of the Aubry-Andre' model to interacting 
fermions is the one with the following Hamiltonian
\bea 
&&H=-\e(\sum_{x} a^+_{x+ 1} a^-_{x}+ a ^+_{x-1} a^-_{x} )+\nn\\
&&u\sum_x
\cos (2\pi(\o x))  a^+_{x} a^-_{x}+U \sum_{x}
a^+_{x} a^-_{x} a^+_{x+1}
a^-_{x+1}
\label{1.1}\eea
where $x\in \mathbb{Z}$ and $a^\pm_{x}$ are fermionic creation or annihilation operators. Such model describes spinless fermions with nearest-neighbor interaction, and it can be mapped, using the Jordan-Wigner transformation, in the XXZ spin chain with quasi-random disorder. The model
\pref{1.1} has been extensively studied numerically, and evidence of MBL phase has been found in 
\cite{H4} and 
\cite{I}. A complete proof of localization of the ground state
of the model \pref{1.1} has been recently 
achieved in
\cite{M}, \cite{M1}, \cite{M2}, by a 
a combination of KAM methods
with nonperturbative  Renormalization Group (RG)
techniques. A proof of a similar property in the random case is still lacking.

The model \pref{1.1} is not really the one realized in cold atoms experiments \cite{B},
which is instead the following
\bea 
&&H=-\e(\sum_{x}\sum_{\s=\uparrow,\downarrow}  a^+_{x+ 1,\s} a^-_{x,\s}+ a ^+_{x-1,\s} a^-_{x,\s} )+\label{1.1aaa}\\
&&u\sum_x\sum_{\s}
\cos (2\pi(\o x))  a^+_{x,\s} a^-_{x,\s}+U \sum_{x}
a^+_{x,\uparrow} a^-_{x,\uparrow} a^+_{x,\downarrow}
a^-_{x,\downarrow}
\nn\eea
Non local dipolar interactions are indeed much more difficult to realize in cold atoms, hence local interactions and spinning fermions are considered.
Aim of this paper is to extend the 
analysis in \cite{M}, \cite{M1}, \cite{M2} to spinning fermions described by the 
model
\pref{1.1aaa}. A crucial difference emerges; while 
in the absence of spin, number-theoretical properties of the frequency and cancellations due to Pauli principle are sufficient the rule out possible relevant effective interactions, in the spinning case
a relevant quartic term is present. At non zero temperature the
number of RG steps is finite and the effective coupling remains
inside the convergence radius if the temperature is not too low.
The large distance exponential decay of the correlations present in the non interacting case, consequence of the single particle  localization, is shown to persist
in the spinning case
only for temperatures greater than a power of many body interaction, while in the spinless case this happens up to zero temperature.

\section{Small divisors and Diophantine numbers}

We consider the
grand canonical 2-point function 
\be <T a^-_{\xx,\s}a^+_{\yy,\s} >={{\rm Tr} e^{-\b (H-\m N)} T a^-_{\xx,\s}a^+_{\yy,\s}
\over{ \rm Tr} e^{-\b (H-\m N)}}
\ee
where $T$ is the time order product,  $\xx=(x_0,x)$,
\be
a_\xx^\pm=e^{(H-\m N) x_0} a^\pm_{\xx,\s} e^{-(H-\m N) x_0}\ee 
and $N=\sum_x\sum_{\s} a^+_{x,\s} a^-_{x,\s}$. 
The chemical potential is chosen equal to  
\be
\m=\cos 2\pi(\o\bar x)
\ee
In the non-interacting case for large $u/\e$ the eigenfunctions are exponentially localized \cite{FS},\cite{Ia}
and as a consequence
$<T a^-_{\xx,\s}a^+_{\yy,\s} >$
decays exponentially for large coordinates difference; our aim is to see if such exponential decay persists in presence of many body interaction.

Already in the non-interacting case, 
\cite{FS},\cite{Ia} one needs to assume not only that the frequency $\o$ is irrational, but that it belongs to a class of irrationals called 
{\it diophantine numbers}, verifying the following property
\be || \o x||\ge C_0
|x|^{-\t}\quad
\forall x\in  \mathbb{Z}/\{0\}
\label{d}\ee with $||.||$ is the norm on the one dimensional torus. Such a property, saying roughly speaking that 
$\o$ is a ``good'' irrational, is not restrictive as 
Diophantine numbers have full measure. \pref{d}
is quite familiar in the theory of dynamical systems,
in KAM theory and generally when one has
to deal with small divisors. We will assume also a 
diophantine condition for the chemical potential
\be ||\o x\pm 2 \o\bar x||
\ge C_0 |x|^{-\t}\quad
\forall x\in  \mathbb{Z}/\{0\}
\label{d1}\ee
Equivalently one can instead consider a quasi-random disorder of the form $\cos (2\pi(\o x+\th))$ and assume a diophantine condition for $||\o x\pm 2 \th||$.

There are two limits in which the correlations can be evaluated easily: one is free fermion limit $\e=1, u=U=0$ and the other is the molecular limit $u=1, \e=U=0$; in this second case
the Hamiltonian is 
$H_0=\sum_x\sum_{\s}
\cos (2\pi(\o x))  a^+_{x,\s} a^-_{x,\s}$ 
with eigenvalues \be \EE=\sum_{x,\s} \cos (2\pi(\o x)) n_{x,\s}\ee
and $n_{x,\s}=0,1$. The 2-point function can be easily computed;
if $\phi_x=\cos (2\pi(\o x))$
\bea &&<T a^-_{\xx,\s} a^+_{\yy,\s}>|_{U=\e=0}=g(\xx,\yy)=\nn\\
&&\d_{x,y}\{ {e^{-(x_0-y_0) (\phi_x-\m)}
\over 1+e^{-\b (\phi_x-\m)}}{\bf I}(x_0-y_0>0)-\\
&&{e^{-(\b+x_0-y_0)(\phi_x-\m)} \over
1+e^{-\b (\phi_x-\m)}}{\bf I}(x_0-y_0\le 0)\}\nn
\eea
The occupation number, defined as the limit $x_0-y_0\to 0^-$, is equal in the limit $\b\to\io$ to $\chi(\phi_x-\m<0)$.
The function $g(\xx,\yy)$ is antiperiodic in $x_0-y_0$ of period $\b$; hence
its Fourier series is of the form
\be
g(\xx,\yy)=\d_{x,y}{1\over\b}\sum_{k_0={2\pi\over\b}(n_0+{1\over 2})}
\hat g(x,k_0)e^{-i k_0 (x_0-y_0)}\ee
with
\be \hat g(x,k_0)=\int_0^\b d\t e^{i \t k_0}{e^{-\t (\phi_x-\m)} \over 1+e^{-\b
 (\phi_x-\m)
}}= {1\over -i k_0+\phi_x-\m}\label{prop} \ee
In the $\e=U=0$ case the 2-point function is vanishing if $x\not=y$. In order to see what happens outside the molecular limit we can expand in $\e, U$ using many body perturbation theory. In the $U=0, \e\not=0$ one could equivalently use that the eigenfunctions of the Hamiltonians are anti-symmetric products
of single particle eigenfunction, and large distance exponential decay of the 2-point function 
should follow from the localization of the single particles eigenstates; when $U\not=0$ there is no knowledge on the eigenstates and series expansion in $\e,U$ for the 2-point function seems the more natural strategy.

While in usual condensed matter problems the lack of convergence is
signaled by divergences in the graphs,
the present case is more subtle; even if finite order by order, 
one can identify graphs
increasing with the order $n$ as a power of $n!$. The presence of such terms is well known in classical mechanics, where are known as {\it small divisors}
; in certain cases they determine divergence of perturbation theory while in others 
resummations still allow to recover convergence. The persistence or not of localization in the present case requires the understanding
of small divisors. If we set $x=x'+\r\bar x$, $\r=\pm$
then 
\be
\hat g(x'+\r \bar x, k_0)\sim {1\over -i k_0\pm v_0  (\o x')_{\rm mod. 1}}
\label{l3}
\ee
and $(\o x')_{\rm mod. 1}$ can be arbitrarily small due to the irrationality of $\o$.
Note the similarity of propagator at the free fermion limit $\e=1,u=U=0$; if $\m=cos p_F$ and $k=k'+\r p_F$ then
\be
{1\over -ik_0+\cos k-\m}\sim {1\over -ik_0+\r v_F k'}
\ee
The points $\pm\bar x$ are Fermi coordinates playing
the role of the Fermi momenta; the similarity between the propagators in the integrable or anti-integrable limit is a manifestation of the well known Aubry-duality.
Looking at \pref{l3}, 
we note that $(\o x')_{\rm mod. 1}$ can be arbitrarily small;
in particular for $||\o x+\o\r\bar x||\le {1\over 4}$ (in the case $\ge {1\over 4}$ the bound is trivial) 
one can write 
 $||\o x+\r\o \bar x||=||\o 2 x+\o \r 2\bar x||/2$ and
 by the diophantine condition $||\o 2 x+\o \r 2\bar x||\ge C_0 |2 x|^{-\t}$ so that  
\be |\hat g(x, k_0)|\le C |x|^\t
\label{l3a}
\ee
The propagator is therefore finite but the above bound
is not sufficient to justify perturbation theory, as one can easily identity contributions apparently preventing convergence.

This can be checked in the
easiest case of the chain graphs of the non interacting theory,  $U=0$ ($\a_i=\pm$), whose value is, if $y=
x+\sum_{i\le n} \a_i$
\be
\e^{n} \hat g(x,k_0)
\prod_{k=1}^{n-1}\hat g(x+\sum_{i\le k} \a_i,k_0)]
\hat g(x+\sum_{i\le n} \a_i,k_0)
\label{gigi}
\ee
By choosing $\a_i=+$ one gets a bound $O(n!^\t)$ apparently spoiling
convergence. Note however that such terms are very similar to the ones appearing in Lindstedt series, expressing the invariant tori in close to integrable systems \cite{GM}; such series are convergent
(despite small 
divisors) due to peculiar compensations and one can guess that a similar mechanism works also here; in the non interacting case we know that this must be true by the results on the single particle spectrum  \cite{FS},\cite{Ia}.
In presence of many body interaction $U\not=0$ 
much more complex graphs appear, namely graphs with  {\it loops}.
Their presence is the signature of an interacting many-body system, and makes the problem genuinely different with 
respect to KAM theory, in which only tree 
graphs appear; here one has to solve a problem with loops and small divisor together.
 
\section{Main result}

We expect that  the interaction causes the renormalization of the chemical potential; it is
then convenient to {\it fix} the interacting chemical potential by choosing properly its bare value. The bare chemical potential is chosen as 
$\m=\cos (2\pi\o\bar x)+\n$, and $\n$ is a function of $\e,U$ which is determined by requiring the the dressed chemical potential is $\cos 2\pi(\o\bar x)$. 
Our main result is the following.
\vskip.3cm
{\rm Theorem.} {\it There exists $\e_0$ such that the 2-point function of the model
(2) verifies,
for $|\e|\le \e_0$, $|U|\le \min(\e_0,\b^{- 2})$,
$\o,\bar x$ verifying \pref{d}, \pref{d1} and for a suitable chosen $\n$, for any $N$
\be|<T a^-_{\xx,\s} a^+_{\yy,\s}>|\le C_N 
e^{- \x|x-y|} {|\log \D|\over 
(1+(\D|x_0-y_0)|)^N}
\label{fon}
\ee
where \be \x=\max(|\log\e|,\b^{-1})\ee
and \be \D=
\max((1+\min(|x|,|y|))^{-\t},\b^{-1})\ee
}
\vskip.3cm
The above result says that the exponential decay ( with rate independent from temperature) in the coordinate of the non interacting case, consequence of single particle  localization,
persists up to a temperature greater than a power of the interaction
; $\e$ can instead be chosen independent from the temperature. 
The decay in imaginary time is much weaker.
The result is  {\it non perturbative}
, in the sense that the 2-point function is written in terms of a renormalized expansion (different from the power series expansion described above)
whose convergence is proved. The estimated radius of convergence $\e_0$ (whose explicit value
follows from the bounds in \cite{M1}) 
is expected to be far from the values where delocalization should happen.
It is however important to know that the radius of convergence is finite to rule out possible non-perturbative effects and the possibility that localization is a metastable phenomenon.
In the spinning case the quartic effective interaction is relevant, and the condition on the temperature is used to ensure that 
it remains small at each iteration of the RG; in the spinless case, on the contrary, the quartic terms are irrelevant and exponential decay can be proved up to zero temperature. Finally we recall that
in the opposite regime, that is $\e=1$ and $u,U$ small the correlations decay instead with a power law, \cite{M3a},\cite{M3}, so that a localization-delocalization 
is present also with a many body interaction.

We present below the main ideas in order to get the above results,
stressing the main new technical problems posed by the presence of spin (for more technical details on the part of the analysis which does not depend from spin we refer to \cite{M},
\cite{M1},\cite{M2}).

\section{Renormalization Group analysis}
 
The 2-point function is the second derivative of the generating function 
\be
e^{W(\phi)}=\int P(d\psi)e^{V(\psi)+(\psi,\phi)}\label{ww}
\ee
with
\bea
&&V=-U\int d\xx \psi^+_{\xx,\uparrow}\psi^-_{\xx,\uparrow} \psi^+_{\xx,\downarrow}\psi^-_{\xx,\downarrow}+\n \sum_\s\int d\xx
\psi^+_{\xx,\s} \psi^-_{\xx,\s}\nn
\\
&&\e\sum_\s\int d\xx (
\psi^+_{\xx,\s} \psi^-_{\xx+{\bf e_1},\s}+\psi^+_{\xx+{\bf e_1},\s}\psi^+_{\xx,\s})
\eea
where $\psi$ are grassmann variables, $\phi$ is the external source, $\int d\xx=\int dx_0\sum_x$,
${\bf e}_1=(0,1)$ and $P(d\psi)$ is the fermionic integration with propagator 
\pref{prop}. We introduce
a cut-off smooth function $\chi_\r(k_0,x)$ which is non vanishing for
$\sqrt{k_0^2+((\o(x-\r \bar x)_{\rm mod. 1})^2)}\le \g$, where $\r=\pm 1$ and $\g>1$
is a suitable constant (to be fixed below); therefore we can write the propagator as 
\be
\hat g(k_0,x)=\hat g^{(u.v.)}(k_0,x)+\sum_{\r=\pm}\hat g_\r(k_0, x)
\ee where $\hat g_\r(k_0, x)={\chi_\r(k_0,x)\over -i k_0+\phi_x-\m}$, $\phi_x=\cos(2\pi(\o x))$ 
and correspondingly
 $\psi_{k_0,x}=\psi^{(u.v.)}_{k_0, x}+\sum_{\r=\pm 1} \psi_{\r, k_0, x}$. This simply says that we are rewriting 
the fermionic field as sum of two independent fields living close to one of the Fermi points, up to a regular field.
We can further decompose \be \hat g_\r(k_0,
 x)=\sum_{h=-\io}^0 \hat g^{(h)}_{\r}( k_0,x)\ee with 
$\hat g^{(h)}_\r(k_0,x)$ similar to $\hat g^{(h)}_{\r}(k_0,x)$ with $\chi$ replaced by $f_h$ 
with
$f_h(k_0,\0 x')$ non vanishing in a region $\sqrt{k_0^2+((\o x')_{\rm mod 1})^2}\sim \g^h$.
Note that at finite temperature $\b<\io$ there is a finite scale $h_\b\sim -\log\b$ so that $f_h=0$ for $h\le h_\b$; that is the number of scales is finite and the temperature acts as in infrared cut-off.

After the integration of the fields $\psi^{(u.v.)}, \psi^{(0)},..,\psi^{(h+1)}$ the generating function has the form
\be
e^{W(\phi)}=\int P(d\psi^{\le h})e^{V^{(h)}(\psi)+B^{(h)}(\psi,\phi)}\label{hh}
\ee
where $P(d\psi^{\le h})$ has propagator $g_\r^{(\le h)}=\sum_{k=-\io}^h g_\r^{(k)}$ and $V^{(h)}(\psi)$
is given by sum of terms
\bea
&&\sum_{x'_1}
\int dx_{0,1}...\int dx_{0,m}
W_m^{(h)}(x'_1, x_{0,m},..,x_{0,m})\nn\\
&&\psi^{\e_1(\le h)}_{\r_1,x_{0,1},x'_1+\r_1\bar x,\s_1}... \psi^{\e_m(\le h)}_{\r_m,x_{0,m},x'_m+\r_m\bar x,\s_m}
\label{ep}\eea
where the kronecker deltas in the propagators imply that a single sum over $x$ is present; the kernels $W_m^{(h)}$ are sum of Feynman diagrams obtained connecting vertices $\e$, $U$
or $\n$ with propagators $g^{(k)}$ with $k>h$. Similarly $B^{(h)}$ is given by a similar expression with the only difference that some of the external lines are associated to $\phi$ fields.
In each of the Feynman diagrams contributing to $W_m^{(h)}$ there is a
{\it tree} of propagators connecting all the vertices and external lines; the coordinates $x_i$, $x_j$ of two external lines
as such that 
\be x_i-x_j=x'_i+\r_i \bar x-x'_j-\r_j \bar x=
\sum^*_\a \d_\a\label{hhh}\ee 
where the sum is over the vertices in the path of the tree connecting $i$ and $j$ and $\d_\a=(0,1,-1)$  is associated to the line connected to the vertex $\a$.

{\it Power counting} arguments says that the theory is {\it non renormalizable}, as the scaling dimension of each term in the effective potential is independent from the number of fields and equal to
$
D=1
$
(they are {\it all relevant}). This says that the size of the kernels apparently increases at each iteration of the Renormalization group, so that one exits from the convergence domain and the RG approach becomes unreliable.
However, dimensional arguments do not take into account
number-theoretic improvements coming from the Diophantine conditions \pref{d}, \pref{d1}. As we will briefly discuss in the following section, a great number of terms, dimensionally
relevant, are indeed irrelevant.

\section{Irrelevance of the non-resonant terms}

We can distinguish in the effective potential two kinds of contributions; the terms such that all the fields have the same coordinates measured from the Fermi points 
, that is $x'_i=x'_1$ for any $i=1,..,n$, are called
{\it resonant terms}.  The others are called {\it non-resonant terms}. In a resonant term 
$\r_i=\r_1$ for any $i=1,..,n$; indeed from \pref{d1}
we see that $2\bar x\not \in Z/\{0\}$; 
as $x_i-x_j\in Z$  and $x'_i=x'_j$ then
$(\r_i \bar x-\r_j \bar x)\in Z$. The resonant terms have therefore the form
\be
\sum_{x'}
\int \prod_{i=1}^n dx_{0,i}
W_m^{(h)}\psi^{\e_1(\le h)}_{\r,x_{0,1},x'+\bar x_{\r_1},\s_1}... \psi^{\e_m(\le h)}_{\r,x_{0,m},x'+\bar x_{\r_m},\s_m}
\label{ep}\ee

We want to show that the non-resonant terms,
even if dimensionally relevant, are indeed irrelevant.
In order to do that we need more accurate bounds.
It is convenient, 
given a Feynman graph, 
to consider a maximally connected subset of lines corresponding 
to propagators 
with scale $h\ge h_v$ with at least a scale $h_v$, and we call it {\it cluster} $v$ (for more details, see \cite{M}); 
the external lines have scale smaller then $h_v$. Therefore to each Feynman graph is associated a hierarchy of 
clusters; inside each cluster $v$ there are $S_v$ maximal clusters, that is clusters   
contained only in the cluster $v$ and not in any smaller one, or trivial clusters given by a single vertex. Each of such $S_v$ clusters are connected 
by a tree of propagators with scale $h_v$; by integrating the propagators over the time, and using that 
\be\int dx_0 |g^{(h)}_\r(x_0, x)|\le C\g^{-h}\quad|g^{(h)}_\r(x_0, x)|\le C\ee we get 
that each graph of order $n$ contributing to $W^{(h)}_m$ is bounded by, if $\bar\e=\max(|\e|,|U|)$ 
\be C^n \bar\e^n 
\prod_v \g^{-h_v(S_v-1)}\label{paz}\ee
where $v$ are the clusters (not end-points) and
$h_v\le 0$. 
If $v_0$ is the largest cluster, $v$ are the clusters
 (without end-points), 
  $\bar v$ the end-points and $R$ or $NR$ the resonant clusters or end-points and $v'$ is the first cluster enclosing $v$, then
$$\prod_{v} \g^{-h_v S_v}=
\prod_{v\not=v_0} 
\g^{-h_{ v'}}\prod
_{\bar v} 
\g^{-h_{\bar v'}}$$
and  
$\prod_{v} \g^{h_v}=\g^{h_{v_0}}
\prod_{v\not= v_0} \g^{h_v}$
so that \pref{paz} can be rewritten as
\be  \g^{h_{v_0}}
\prod_{v\not=v_0} 
\g^{-D (h_{ v'}-h_v)}\prod_{\bar v } 
\g^{-h_{ \bar v'}}\label{xak1}
\ee
with $D=1$ is the scaling dimension.
In getting \pref{paz}, 
we have however not used a crucial property implied by the Diophantine condition: namely
that if the denominators associated to the external lines have the same size but different coordinates, the difference of coordinates must be 
large and therefore there is a decaying factor associated to a large number of $\e, U$. 

Indeed in the non resonant terms
there are least two external fields with coordinate $x'_1, x'_2$ with $x'_1\not= x'_2$; 
if $v'$ is the cluster containing $v$
\bea
&&2 \g^{h_{ v'}}\ge ||(\o x'_{1})||+ ||(\o x'_{2})||\ge 
||\o(x'_{1}-x'_{2})||=
\\
&&||(\r_2\bar x-\r_1\bar x)\o
+\o \sum^*_\a \d_\a ||\nn
\eea
If $\r_{1}=\r_{2}$ we use \pref{d}
and if $\r_{1}=-\r_{2}$ we use 
\pref{d1}; we get
\be
2 \g^{h_{ v'}}\ge {C_0\over |\sum^*_\a \d_\a |^{\t}}\ee so that 
$|\sum^*_\a \d_\a | \ge \tilde C \g^{-h_{ v'}/\t}$. 
This says that the non-resonant terms have external fields 
with a difference of coordinates which is quite large, and
as $|\sum^*_\a \d_\a |$ is surely greater than the number of $\e,U$ vertices in the cluster $v$, this
produces a gain factor. Writing $\bar\e=\max(|\e|,|U|)$ 
\be
\bar\e=\prod_{h=-\io}^0 \bar\e^{2^{h-1}}
\ee
we can associate a factor $\bar\e^{2^{h_v-1}}$ for each end-point enclosed in the cluster $v$; as 
$|\sum^*_\a \d_\a |$ is surely smaller that the number of end-points and using that  $\g^{1\over\t}/2>1$ and $\bar\e$ small
\be
\bar \e^{n\over 2}\le \prod_v \bar \e^{C 2^{h_v}\g^{-h_v/\t}
S^{NR}_v}\le \prod_{v\in NR}\g^{
2 h_{v'}}\prod_{\bar v \in NR}\g^{
2 h_{\bar v'}}
\label{xak}
\ee
where $S^{NR}_v$ are the non resonant clusters contained in $v$;
that is a decay factor is associated to each non resonant cluster.

\section{Renormalization of the resonant terms}

A gain is found for the non-resonant terms due to the Diophantine condition which makes them irrelevant.
In the resonant terms
with more than $6$ external fields, there are at least two couple of fields with the same $\e,\r,\s$ (remember that $\r$ in the resonant term is the same) and we can write
\be
\psi^\e_{x',x_{0,1},\r,\s}\psi^\e_{x',x_{0,2},\r,\s}=
\psi^\e_{x',x_{0,1},\r,\s}(\psi^\e_{x',x_{0,2},\r,\s}-\psi^\e_{x',x_{0,1},\r,\s})
\ee
and \be
\psi^\e_{x',x_{0,2},\r,\s}-\psi^\e_{x',x_{0,1},\r,\s}=(x_{0,2}-x_{0,1})\int_0^1 dt \partial \psi^\e_{x',x_{0}(t),\r,\s}
\ee
with $x_{0}(t)=x_{0,1}+t(x_{0,2}-x_{0,1})$.
The derivative produces an extra 
$\g^{h_{v'}}$ and the factor $(x_{0,2}-x_{0,1})$
an extra $\g^{-h_{v}}$; as there are at least two 
of such monomials one gets at least a factor 
$\g^{2(h_{v'}-h_v)}$ making the dimension negative.
The same factor is obtained for similar reasons in the quartic terms with the same spin.
We have therefore to renormalize the terms with $2$ and $4$ external lines. 
The integration of the field $\psi^h$ is done writing (in the $\phi=0$ case for definiteness)
\be
\int P(d\psi^{\le h})e^{\LL V^{(h)}(\psi^{\le h})+\RR V^{(h)}(\psi^{\le h})}
\ee
with $\RR=1-\LL$ and $\LL$ is the localization operator
acting on the terms with 2 and 4 external lines in the following way. If $\hat W^h_2(k_0,x')$ is the kernel of the terms with 2 external lines, $\LL$ is defined as
\be
\LL \hat W^h_2(k_0,x')=\hat W^h_2(0,0)+k_0 \partial_0 \hat W^h_2(0,0)+(\o x')\tilde\partial\ W^h_2(0,0)
\ee
where $\tilde\partial \hat W^h_2(k_0,x')={\tilde W^h_2(k_0,x')-
\tilde W^h_2(k_0,0)\over(\o x')}$. Again the action of $\RR=1-\LL$ produces a gain $\g^{2(h_{v'}-h_v)}$, using also that 
$(\o x')^2
\sim \g^{2h_{v'}}$ for the compact support properties of the lines external to the cluster $v$, while 
$\tilde \partial^2 \hat W^h_2(k_0,x')$ has an extra 
$\g^{-2h_v}$.
%
Finally the action of $\LL$ on the kernels with four external fields 
$ \hat W^h_2(\underline k_{0};
x')$ is 
\bea
&&\LL\hat W^h_2(\underline k_{0};x')=
\hat W^h_2(\underline 0;0)+\\
&&(\o x')\tilde\partial \hat W^h_2(\underline 0;0)+\sum_{i=1}^4 k_{0,i}\partial_{0,i}\hat W^h_2(\underline 0;0)\nn
\eea
The $\RR$ operation produces therefore an extra factor 
\be \prod_{v\in R}\g^{2(h_{v'}-h_v)}\ee By construction \be \LL\VV^h=\LL\VV_a^h+\LL\VV_b^h\ee
with $\LL\VV_a^h$ contains the relevant terms
\bea
&&\LL\VV_a^h=\g^h\n_h\sum_{\r,\s}\int d\xx \psi^+_{\xx,\r,\s}
 \psi^-_{\xx,\r,\s}+\nn\\
&&\g^h\l_h\sum_\r\int d\xx \psi^+_{\xx,\r,\uparrow}
\psi^-_{\xx,\r,\uparrow}\psi^+_{\xx,\r,\downarrow}
\psi^-_{\xx,\r,\downarrow}
\eea
with $\l_0=U$, and $\LL\VV_b^h$
contains the marginal terms, that is monomials like $
\psi^+\partial\psi^-$ and $\psi^+\partial\psi^-\psi^+\psi^-$
times the corresponding running coupling constants. 
Note that $\n_h$ is independent from $\r$ by parity. 

In the spinless case the quartic relevant and marginal term
in $\LL\VV_a^h$ would be vanishing by Pauli principle; 
with spin they are non vanishing and with a non trivial flow.
We write then  \be P(d\psi^{\le h})= P(d\psi^{\le h-1}) P(d\psi^{h})\ee  and we
integrate the field $\psi^h$ 
so that procedure can be iterated.

Operating in this way we get an expansion in terms of running coupling constants $\vec v_h=(\l_h, \n_h, \z_h)$ where $\z_h$ are the marginal couplings in $\LL\VV_b^h$; therefore
$W^{(h)}_m$ can be bounded by, instead  by \pref{paz}, by,
if $\bar\e_h=\max (|\l_h|,|\n_h|,|\z_h|,|\e|)$
\bea
&& C^n \bar\e_h^n \prod_v \g^{-h_v(S_v-1)}]\\
&&[\prod_{v\in R}\g^{2(h_{v'}-h_v)}]
[\prod_{\bar v\in R} \g^{h_{\bar v'}}] [\prod_v \e^{C 2^{h_v}\g^{-h_v/\t}
S^{NR}_v}]\label{zaz11}\nn
\eea
and using \pref{xak1} and \pref{xak}
we get that (42) can be written as
\be C^n \bar\e_h^n
 \g^{h_{v_0}}[\prod_{v}\g^{(h_{v'}-h_v)}][\prod_{\bar v\in NR} \g^{h_{\bar v'}}]
\ee
As $h_{v'}-h_v\le 0$ it is possible over the scales $h_v$ obtaining a
$O(C^n\bar\e_h^n)$ bound from which convergence follows provided that $\e_h$ is not too large. Note that the above bound is valid for the sum of all Feynman graph of order $n$, by using determinant bounds for fermionic expectations, see  
\cite{M}, for details. Moreover one needs to sum over the number of external lines of each cluster and a corresponding decaying factor is necessary. It is
obtained by considering a tree of propagators  connecting the external lines with the same $\r,\s$ index, and noting that, for any couple of lines, either the coordinates are the same, hence the couple of external lines can be rewritten as (34) and 
the decay factor is obtained by (35), or are different and the decay factor is from (31); see \cite{M2} \S 3.F for more details.

\section{The flow of the running coupling constants}

We need to show that the running coupling constants remain small iterating the Renormalization Group flow.
The relevant quadratic coupling verify the equation
\be
\n_{h-1}=\g\n_h+\b^h_2\label{bb1}
\ee
where $\b^h_2=\g^{-h} W_2^h$; we can write 
$\n_{h}=\g^{-h} (\n+\sum_{k=0}^{h+1}
\g^{k-1}\b^k_2)$  and by choosing the counterterm
$\n$ as $\n=-\sum_{k=0}^{h_\b}
\g^{k-1}\b^k_2$ we get that $\n_h$ is bounded uniformly in $h$.
The relevant quartic effective coupling verifies
\be
\l_{h-1}=\g\l_h+\b^h_4\label{bb}
\ee
where $\b^h_4=\g^{-h} W_4^h$; we can write 
$\b^h_4=\b^h_a+\b^h_b$ where $\b^h_a$ contains no contributions from irrelevant vertices, and at least 2 
quartic end-points are present by the compact support properties 
of the propagator. In $\b^h_b$ there is at least an irrelevant end-point and a quartic one, so that
${\l_{h-1}\over \l_h}=\g( 1+O(\l_h)+O(\e\g^h))$ from which
\be |\l_h|\le \g^{-h {5\over 4}} |U|\le |U|^{3\over 8}\ee Similarly the marginal quartic terms $\tilde\l_h$ verify 
$\tilde\l_{h-1}=\tilde\l_h+\tilde\b^h_4$ with $\tilde\l_0=O(U(U+\e))$ and 
$\tilde\b^h_4$ verifies the same bounds than $\tilde\b^h_4$ (no contributions with a quartic and a quadratic running coupling constant) so that $|\tilde \l_h|\le |U|^{3\over 8}$. Finally to the beta function of the marginal quadratic terms $\tilde\z_h$ no contributions depending only on $\n_h,\tilde\z_h$ are present by the compact support properties of the propagator; therefore 
it can be written as sum of the two terms, one depending from at least an irrelevant end-points, which is $O(\e \g^h)$, and the other with at least a quartic running coupling constant, which 
is $O(|U|^{3\over 8})$, so that the final size is 
$O(\log |U |U|^{3\over 8})+O(\e)$. In the spinless case there are no quartic couplings; the beta function of the quadratic marginal coupling is 
$O(\e \g^h)$ so that the boundedness of the flow follows, wth a proper choice of $\n$, without any finite temperature condition.

In conclusion, we get a convergent expansion for the 2-point function in terms of running coupling constants; if $x$, $y$ are the coordinates of the external lines, there are at least a number $\ge |x-y|$ of $\e$ factors, so exponential decay in the coordinates follows.

\section{Conclusions}

We have analyzed the role of the interaction
in a system of fermions with a quasi-random disorder, in the regime where localization is present in the single particle spectrum. The results are non-perturbative, in the sense that 
they are written in terms of a renormalized  expansion in running coupling constant with a finite disk of convergence.
A clear difference between spinless and spinning fermions emerges; in the first case there are no relevant effective interactions while 
in presence of the spin an additional relevant
quartic term is present in the RG flow. Such effective coupling tends to increase iterating the RG; to keep it inside the estimated convergence radius (and establish exponential decay of correlations)
we impose a finite temperature condition, while in the spinless case convergence is up to zero temperature.

The presence of an additional effective coupling, due to spin, naturally indicates the emergence of new phenomena at lower temperatures. This is confirmed
by recent numerical simulations 
\cite{P} in which is found a different behavior in the spinless and spinning case, and in particular different trasport properties 
for the charge or spin densities in the spinning case. The extra coupling we found in the spinning case
could flow to a non trivial fixed point with
different large distance behavior in spin or charge densities, as it
happens in 
weakly interacting one dimensional fermions \cite{S}; a rigorous confirmation of this scenario requires the extension of the methods used here to deal with strong coupling problems.


\begin{thebibliography}{0}
 \bibitem{A}%
 P. W. Anderson
{\it 
Absence of Diffusion in Certain Random Lattices.} 
Phys. Rev. 109, 1492 (1958)
  
\bibitem{PH1}%
V. Oganesyan, D. A. Huse:{\it 
Localization of interacting fermions at high temperature.}
Phys. Rev. B 75, 155111 (2007)


\bibitem{PH4}%
A. Pal, D.A. Huse:  {\it 
Many-body localization phase transition. }
Phys. Rev. B 
82,  174411 (2010)


\bibitem{NH}%
R. Nandkshore, D. Huse:
{\it Many body localization and thermalization in quantum statistical mechanics}
Annual Review of Condensed Matter Physics, Vol. 6: 15-38 (2015)

\bibitem{HL}%
S. Goldstein, D. A. Huse, J. L. Lebowitz, R. Tumulka: 
{\it Thermal equilibrium of a macroscopic quantum system in a pure state}
Phys. Rev. Lett. 115, 100402 (2015) 

\bibitem{FA} L.Fleishmann, P.W. Anderson
{\it Interactions and the Anderson transition}
Phys. Rev B 21, 2366 (1980)

\bibitem{GGG}
Gornyi, I., Mirlin, A., and Polyakov, D. 
{\it Interacting electrons in disordered wires.}
Phys. Rev. Lett. 95 (2005), 206603.





\bibitem{Ba}%
 D.M. Basko, I. Alteiner , B. L. Altshuler: {\it 
Metal-insulator transition in a weakly interacting many-electron system with localized single-particle states.} 
Ann. Phys. 321, 1126 (2006)

\bibitem{Ba1}%
M. Serbyn, Z. Papic, A. Abanin: {\it
Local Conservation Laws and the Structure of the Many-Body Localized States} 
Phys.
Rev. Lett. 111, 127201 (2013).

\bibitem{Ba2}%
V. Ros, M. Mueller, A. Scardicchio: {\it
Integrals of motion in the Many-Body localized phase
 } Nuclear Physics
B
891
, 420 (2015).

\bibitem{Ro}%
W. De Roeck, F. Huveneers, M. Müller, and M. Schiulaz; 
{\it Absence of Many-Body Mobility Edges} Phys. Rev. B 93, 014203 (2016).

\bibitem{Ga}%
G. Gallavotti: {\it Quasi integrable mechanical systems} , ed. K.Osterwalder and R.Stora, Les Houches, XLIII, 1984, Phenome'nes Critiques, Syste'mes aleatories, The'ories de jauge, Elsevier Science, 1986, p.539--624. 


\bibitem{loc4}%
J. Imbrie: {\it 
On Many-Body Localization for Quantum Spin
Chains.} 
Jour. Stat. Phys. 163:998-1048 (2016)

\bibitem{loc4a}%
J. Imbrie:
{\it Diagonalization and Many-Body Localization for a Disordered Quantum Spin Chain.} 
Phys. Rev. Lett. 117, 027201 (2016) 

\bibitem{S1}%
E. Hamza, R. Sims,  and G. Stolz, 
{\it Dynamical Localization in Disordered Quantum Spin Systems
}
Commun. Math. Phys.
315
, 215 (2012).

\bibitem{S2}%
R. Seiringer, S. Warzel: 
{\it
Decay of correlations and absence of superfluidity in the disordered Tonks–Girardeau gas}
New Journal of Physics, Volume 18, March 2016 




\bibitem{B}%
M Schreiber 
S. Hodgman
P Bordia, 
H P. Lüschen,
M. H. Fischer, 
R. Vosk, 
E. Altman, 
U. Schneider, 
I. Bloch:
{\it Observation of many-body localization of interacting fermions in a quasi random optical lattice}
Science  
Vol. 349, Issue 6250, pp. 842-845
2015

\bibitem{H4}%
S. Iyer, V. Oganesyan, G. Refael, D. A. Huse: {\it 
Many-body localization in a quasiperiodic system.}
 Phys. Rev. B 87, 134202 (2013)



\bibitem{AA}%
S. Aubry and G. Andr\'e: {\it Analyticity breaking and Anderson localization in incommensurate lattices} 
Ann. Israel Phys. Soc 3, 133 (1980).

\bibitem{B1}
P. Bordia, H Lüschen, S. Hodgman, M. Schreiber, I. Bloch, U. Schneider:
{\it  
Coupling Identical 1D Many-Body Localized Systems}
Phys. Rev. Lett. 116, 140401 (2016) 





\bibitem{FS}%
J. Froehlich, T. Spencer, T. Wittwer: {\it 
Localization for a class of one-dimensional quasi-periodic Schrödinger operators},
Comm. Math. Phys.132,1, 5 (1990)




\bibitem{Ia}%
S. Jitomirskaya: 
{\it  Metal-insulator transition for the almost Mathieu operator.}
Ann. of Math. (2)
150 (1999), no. 3, 1159 1175.



\bibitem{I}%
F. Iemini, A. Russomanno, D. Rossini, A. Scardicchio, R. Fazio
{\it  Signatures of many-body localisation in the dynamics of two-sites entanglement.} arXiv:1608.08901

\bibitem{M}%
V. Mastropietro:  {\it Localization in the ground state of an interacting quasi-periodic fermionic chain}
Comm. Math. Phys. 342, 1, 217-250 (2016)

\bibitem{M1}%
V. Mastropietro {\it Localization of interacting fermions in the Aubry-Andr\'e model}
Phys. Rev. Lett. 115, 180401 (2015)

\bibitem{M2}%
V. Mastropietro {\it Localization in interacting fermionic chains with quasi-random disorder}
V Mastropietro
arXiv:1604.08264 (Comm.  Math. Phys. DOI
10.1007/s00220-016-2773-x)


\bibitem{M3a}%
V.Mastropietro: {\it 
Small denominators and anomalous behaviour in the
incommensurate Hubbard-Holstein model}.
Comm. Math. Phys. 201, 81 (1999)


\bibitem{M3}%
V. Mastropietro 
{\it Dense gaps in the interacting Aubry-Andre' model}
Phys. Rev. B 93, 245154 (2016)


\bibitem{GM} G.Gentile, V.Mastropietro: {\it 
Methods for the analysis of the Lindstedt series for KAM tori and
renormalizability in classical mechanics. A review with some applications}
Rev. Math. Phys. 8, 3, 393 (1996).

\bibitem{P}%
 P. Prelovsek, O.S. Barisic, M. Znidaric
{\it 
Absence of full many body localization in disordered Hubbard chain} 
arxiv1610.02267



\bibitem{S} J. Solyom: {\it
The Fermi gas model of one-dimensional conductors}
 Advances in Physics
28, 2010-303 (1979) 
  
\end{thebibliography}
\end{document}